# Collapse of the quantum wavefunction and Welcher-Weg (WW) experiments

Y.Ben-Aryeh

Physics Department, Technion-Israel Institute of Technology, Haifa, 32000 Israel

e-mail: phr65yb@ph.technion.ac.il

**Absstract**

The 'collapse' of the wave function in a general measuring process is analyzed by a pure quantum mechanical (QM) approach. The problem of the delayed choice and Welcher-Weg (WW) experiments is analyzed for Mach-Zehnder (MZ) interferometer. The WW effect is related to complementarity principle and orthogonality of wavefunctions although it produces small momentum changes between the electromagnetic (EM) field and the beam-splitters (BS's). For QM states for which we have a superposition of states and interference effects there is no reality before measurement, but by excluding such effects the system gets a real meaning already before measurement.

# 1. Introduction

Fundamental issues in physics are the relations between Quantum Mechanical (QM ) and Classical effects. While QM processes are produced by reversible unitary processes in classical theories the effects are quite often irreversible. The interaction between a microscopic QM system and a measuring macroscopic apparatus leads to the 'collapse' of the wavefunctions [1,2] which is basicly an irreversible process, and its interpretation by pure QM might raise some problems. A fundamental problem in QM arises by the claim that reality is obtained only after measurement [3]. Especially interesting in this connection are the delayed–choice experiments since as Wheeler has commented on this problem: "the most startling is that seen in the delayed –choice experiment. It reaches back into the past in apparent opposition to the normal order of time" [4]. I would like to treat in the present paper these problems using only pure QM approach, and compare my analysis with those of other authors. Let me explain further these problems and the methods which I choose for their solutions.

The basic problem in the interpretation of quantum measurement can be explained as follows: If the state is not an eigenstate of the observable, no determinate value is attributed to the observable. However, the measurement is described by a projection postulate [1,2] which characterizes the 'collapse' of the quantum state of the system into an eigenstate of the measured observable. According to "orthodox quantum mechanics (QM)" the observer gets a determinate value as the outcome of measurement [3]. Although this assumption , referred to as a collapse, seems to be in agreement with the experimental measurements, conceptually, it raises the problem of nonunitary evolution, which is not included in pure QM. Wheeler explained Bohr attitude using a single simple sentence: "No elementary phenomenon is a phenomenon until it is registered (observed) phenomenon" [4]. Conceptually it raises the question if the quantum world has an objective meaning before measurement. I would like to analyze in the present paper the interaction between the microscopic quantum system and the macroscopic apparatus of measurement. In such interaction entangled states are produced in which the system wavefunctions are correlated with a certain degree of freedom of the apparatus, depending on the chosen operator of measurement. This interaction is a completely unitary process. The 'real' measurement is made on the macroscopic degree of freedom of the apparatus giving information on the quantum state due to correlation. After measurement, the macroscopic degree of freedom is excluded by a partial trace operation over the entangled states. There are two options for such trace operations by which the system wavefunction collapses which are analyzed both mathematically and physically in Section 2.

Related to these measurement effects we can examine Wheeler [4] proposal of "delayed-choice" Gedanken Experiment in interferometer in which the choice of which property will be observed is made after the photon has already passed the first beam splitter $BS_{inp}$. "Thus one decides whether the photon shall have come by one route or by both routes after it has already done its travel" [4]. There is a lot of literature verifying experimentally the delayed choice phenomena. (see e.g. [5]). The distinction between classical and quantum states can resolve this problem. The idea that the photon has already done its travel is not correct. As long as we have superposition of quantum states, entanglement and interference effects there is not any reality before measurement. This is in agreement with Bohr attitude [4]. However, when we add the Welcher-Weg (WW) detector or omitting in an interferometer the second beam splitter $BS_{out}$, this leads to orthogonality relation between the two routes of the optical system (i.e. eliminating the interference terms) and then the system gets a classical meaning as the 'which-way' the photon traveled is fixed already before measurement.

The WW phenomena have raised conceptual problems. Scully and his collegues [6,7] have claimed that the WW phenomena follows from the complementarity principle. They explained the complementary principle as follows : "We say that two observables are 'complementary' if precise knowledge of one of them implies that all possible outcomes of measuring the other one are equally probable". They have given [6,7] examples illustrating this property. In the WW experiments complementarity means that simultaneous observation of wave and particle behaviour is prohibited. Storey et al. [8] claimed that "interference is lost by the transfer of momentum to the particle whose path is determined ". Englert, Scully and Walther have responded [9] in the negative claiming that complementarity must be accepted as independent principle in quantum mechanics, rather then a consequence of the position-momentum uncertainty relation. Storey et al. by making certain calculations remained in their opinion [10]: "As momentum is conserved in every interaction in quantum mechanics, any process that changes one pattern into another must involve momentum transfer". Experiments [11,12] have shown that the WW apparatus that "determines" which way might not include any momentum transfer, and this is in contradiction with the claims made by storey et al [8,10]. Englert [13] has found a way to quantify which way using a certain inequality but as pointed by him :"The derivation of the inquality does not make use of Heisenberg 's uncertainty relation in any form".

Sculman [14] has analyzed the "two-slit experiment" showing the determination of the slit transversed destroys interference. As explained by him " the destruction of interference comes about because the detected portion of the wavefunction is orthogonal to that which comes out from the other slit. The wavefunction passing through becomes entangled with degrees of freedom of the detector". While such claims are in agreement with our previous analysis [15,16] there is a certain difference. As pointed by him you "must take into account the larger Hilbert space, the detector as well". But in the total Hilbert space one needs to take into account the recoil of the BS's in the interferometers or the wall at 2-slit experiment. As shown in our previous papers [15,16 ] the measurement in one location (the WW detector) influences the wavefunction in other locations and fixes momentum at that second place, exactly like that of EPR. It is true that for bodies which are macroscopic the change in their wavefunctions is quite small and they cannot be considerd as the cause for the WW effect [6-7,9]. But, as many photons are involved in the interference effects Storey et al. are right in their claim that momentum transfers should be involved in the WW experiments. I give in Section 3 detailed calculations for such momentum transfers in Mach-Zehnder (MZ) interferometer using the methods developed in our previous publications [15-16], showing that momentum transfers indeed occur but that they are the result of the orthogonality relations [14], or equivalently of the complementarity properties [6,7,9,11]. In Section 4 the present results and conclusions are summarized.

**2. Collapse of the wavefunction in the measurement process**

There are two stages in the measurement process: 1) In the first stage, the interaction of the measuring apparatus with the quantum system leads to correlations between the quantum states of the system and a certain degree of freedom of the measuring apparatus. This stage is obtained by a completely unitary process leading to the entangled quantum state

$$\left|\psi^{(n)}\right\rangle = \sum_j a_j^{(n)} \left|\xi_j^{(n)}\right\rangle \left|\eta_j^{(n)}\right\rangle \tag{1}$$

where $\left|\xi_j^{(n)}\right\rangle$ are the eigenstates of the quantum system 'measuring' operator $\hat{O}_\xi^{(n)}$ with eigenvalues $\varepsilon_j^{(n)}$, correspondingly, $a_j^{(n)}$ are the amplitudes of the corresponding entangled states multiplications $\left|\xi_j^{(n)}\right\rangle\left|\eta_j^{(n)}\right\rangle$, and $\left|\eta_j^{(n)}\right\rangle$ are the 'correlated' measuring device wavefunctions. We assume that the apparatus quantum states $\left|\eta_j^{(n)}\right\rangle$ are orthonormal. One should notice that we have chosen here a special operator $\hat{O}_\xi^{(n)}$ referred here by the superscript ($n$) leading to a special form of the entanglement given by Eq. (1). Choosing another measuring operator all the entanglement parameters will vary correspondingly. We express the entangled state of Eq. (1) by the density operator

$$\left|\psi^{(n)}\right\rangle\left\langle\psi^{(n)}\right| = \left\{\sum_j a_j^{(n)} \left|\xi_j^{(n)}\right\rangle\left|\eta_j^{(n)}\right\rangle\right\}\left\{\sum_k a_k^{*(n)} \left\langle\xi_k^{(n)}\right|\left\langle\eta_k^{(n)}\right|\right\} \tag{2}$$

As pointed by Peres [17] the interaction between the system and the apparatus produces entangled state due to <u>a unitary interaction</u> between the microscopic particles and the quantum wavefunctions of apparatus: "- it cannot be anything else, if quantum theory is correct". However, the mechanism of entanglement described by the above analysis is different from that of [17].

2) In the second stage of measurement the pure density operator of Eq. (2) 'collapses'. Let us describe first the 'mathematical' mechanism which is behind the process of collapse and then explain its physical meaning and implications. We have two options to do that :

a) By tracing over all macroscopic correlated states and assuming

$$\left\langle\eta_j|\eta_k\right\rangle = \begin{cases} 1 & for \quad j = k \\ 0 & for \quad j \neq k \end{cases}, \tag{3}$$

we get

$$\sum_l \left\langle\eta_l^{(n)}\left|\psi^{(n)}\right\rangle\left\langle\psi^{(n)}\right|\eta_l^{(n)}\right\rangle = \sum_l \left|a_l^{(n)}\right|^2 \left|\xi_l^{(n)}\right\rangle\left\langle\xi_l^{(n)}\right| . \tag{4}$$

We find in this case that after measurement the quantum state which has been described originally as a superposition of quantum states is converted to a "statistical state". The physical meaning behind this trace operation is that we can measure the 'probabilities' for getting the apparatus quantum states $\left|\eta_l^{(n)}\right\rangle$ which are equal to probabilities of having the correlated quantum system $|\xi_l^{(n)}\rangle$ as given by Eq.(4). The physical conclusion is that we can predict the outcome obtained by the apparatus of measurement only with a certain probability. In order to realize the complete statistics coming out from the measurements we have to repeat the preparation of the system under the same conditions and average the results over the outcomes so that the probability for getting eigenvalue $\varepsilon_l^{(n)}$ corresponding to the eigenstates $|\xi_l^{(n)}\rangle$ which are the eigenstates of a certain operator of measurement $\hat{O}_\xi^{(n)}$ is given by $|a_l^{(n)}|^2$.

b) If, however, for one sample we have the result of a measurement over the macroscopic degree of freedom corresponding to the eigenvalue $\varepsilon_l^{(n)}$ (or by elimination all results which do not give $\varepsilon_l^{(n)}$) then we have

$$\frac{\langle \eta_l^{(n)}|\psi^{(n)}\rangle\langle\psi^{(n)}|\eta_l^{(n)}\rangle}{|a_l^{(n)}|^2} = |\xi_l^{(n)}\rangle\langle\xi_l^{(n)}| \qquad . \tag{5}$$

In Eq. (5) the original quantum state has been projected into a special quantum state which after normalization is given by $|\xi_l^{(n)}\rangle$. Such state will start its new time evolution as a pure state.

While such results are in agreement with the ordinary description of "the collapse of the wavefunction" [2,18], we find that the essential source for such 'collapse' comes from the entanglement of the quantum system wavefunctions with those of the apparatus, obtaining the information from the measurements of the apparatus system and reducing the density operator by the trace operation over macroscopic states, excluding them.

Let us illustrate the present approach to the collapse of the wavefunctions by analyzing Stern-Gerlach experiments for 1/2 spin system using it as a prototype simple case of measurement. The entangled state is produced by a magnetic field gradient where the electron beam with 1/2 spin is splitted into two beams with different directions. Without going into all the details of measurement of this system [19] we note that by the above first stage of measurement we get the entangled state

$$|\psi\rangle = a_1^{(n)}|+\rangle^{(n)}|\phi_+\rangle^{(n)} + a_2^{(n)}|-\rangle^{(n)}|\phi_-\rangle^{(n)} \qquad , \tag{6}$$

where the superscript (n) refers to the axis of quantization, $|+\rangle^{(n)}$ and $|-\rangle^{(n)}$ correspond to the spin states with spin eigenvalues of $1/2$ and $-1/2$, $|\phi_+\rangle^{(n)}$ and $\left|\phi_{-}\right\rangle^{(n)}$ are the apparatus states representing the directions of the beams which are entangled with the system spin states, and $a_1^{(n)}$ and $a_2^{(n)}$ are the amplitudes for the entangled states, respectively. The entangled state of Eq. (6) corresponds to Eq. (1) as a special simple case. Let us explain the two physical options of measurements: a) The detectors described by the macroccopic wave functions $\left|\phi_{+}\right\rangle^{(n)}$ and $\left|\phi_{-}\right\rangle^{(n)}$ measure the number of particles $m$ and $n$ going in each direction, respectively, so that $m/(m+n) = \left|a_1^{(n)}\right|^2$, $n/(m+n) = \left|a_2^{(n)}\right|^2$. Then the reduced density operator is given by Eq. (4). b) In another possibility: After separating the spin states so that they are going in different directions let us assume that we put the detectors only in one direction, and excluding those spin particles going in this direction by detecting them. Then the spin states which are going in the other direction and have not been detected can be considered as pure states after normalizing with well defined spin eigenvalue $\varepsilon^{(n)}$ ($\left|+\right\rangle^{(n)}$ with eigenvalue 1/2 or $\left|-\right\rangle^{(n)}$ with eigenvalue -1/2).

It should be pointed out that the above analysis has treated only the "von Neumann measurements" (often called as "measurements of the first kind"). There is a more general type of measurement known as "a positive operator value measure (POVM)" [18]. The fundamental issue of irreversibility is the same for all these measurements [17].

## 3. Welcher-Weg (WW) experiments, complementarity, orthogonality, and momentum transfers in Mach-Zehnder (MZ) interferometer

The WW effects in MZ interferometer can be related to the collapse of the wavefunction following our previous analysis [15,16]. It will be shown here that the WW detector leads to a transfer of momentum between the electromagnetic field and the BS's although there might not be any change of momentum of the WW detector itself.

Assuming creation and annihilation operators $\hat{a}_1^{\dagger}$ and $\hat{a}_2^{\dagger}$ inserted into the input ports of the MZ interferometer, then the transformation of the first beam-splitter ($BS_{inp}$) is given by [16]:

$$\begin{pmatrix} \hat{a}_1^\dagger \\ \hat{a}_2^\dagger \end{pmatrix} = \begin{pmatrix} t_1^* & -r_1 \exp(-\delta\vec{p}\cdot\vec{\nabla}_1) \\ r_1^* \exp(\delta\vec{p}\cdot\vec{\nabla}_1) & t_1 \end{pmatrix} \begin{pmatrix} \hat{b}_1^\dagger \\ \hat{b}_2^\dagger \end{pmatrix} \quad . \tag{7}$$

Here $t_1$ and $r_1$ are the transmission and reflection coefficient of $BS_{inp}$, respectively. $\delta\vec{p}$ is the momentum change of a photon reflected from the $BS_{inp}$ which has been transferred into momentum changes of $BS_{inp}$ with opposite direction, in agreement with conservation of momentum. The unitary matrix transformation of Eq. (7) includes the operator $\exp(-\delta\vec{p}\cdot\vec{\nabla}_1)$ of momentum translation operating on the macroscopic wavefunction of $BS_{inp}$. We get, however [15]:

$$\langle BS_{inp} | \exp(-\delta\vec{p}\cdot\vec{\nabla}_1) | BS_{inp} \rangle \simeq 1 \quad . \tag{8}$$

Here $|BS_{inp}\rangle$ is the macroscopic wavefunction of $BS_{inp}$. The result (8) follows from the fact the photon momentum is very small in comparison to the uncertainty in the momentum of macroscopic object. This is also the reason for neglecting the momentum translation operators of Eq. (7) in conventional treatment of MZ interferometer. We will, however, keep the momentum translation operators in our analysis as it can later resolve the controversy between Scully et al [6-7,9,11] and Storey et al. [8,10] about momentum transfers in WW experiments.

Omitting first the effect of WW detector we have two more transformations. The difference in phase for the two modes entering $BS_{out}$ due to a difference in optical path, can be represented by [16]:

$$\begin{pmatrix} \hat{b}_1^\dagger \\ \hat{b}_2^\dagger \end{pmatrix} = \begin{pmatrix} \exp(i\phi) & 0 \\ 0 & 1 \end{pmatrix} \begin{pmatrix} \tilde{b}_1^\dagger \\ \tilde{b}_2^\dagger \end{pmatrix} \quad . \tag{9}$$

Here $\tilde{b}_1^\dagger$ and $\tilde{b}_2^\dagger$ are the creation mode operators entering $BS_{out}$. A similar transformation to that of Eq. (7) is obtained for $BS_{out}$ [16]:

$$\begin{pmatrix} \tilde{b}_1^\dagger \\ \tilde{b}_2^\dagger \end{pmatrix} = \begin{pmatrix} t_2^* & -r_2 \exp(\delta\vec{p}\cdot\vec{\nabla}_2) \\ r_2^* \exp(-\delta\vec{p}\cdot\vec{\nabla}_2) & t_2 \end{pmatrix} \begin{pmatrix} \hat{a}_3^\dagger \\ \hat{a}_4^\dagger \end{pmatrix} \quad . \tag{10}$$

Combining the transformations (7) (9) and (10) we obtain our MZ transformation expressing the input creation operators $\hat{a}_1^\dagger$ and $\hat{a}_2^\dagger$ as function of the output creation operators $\hat{a}_3^\dagger$ and $\hat{a}_4^\dagger$

$$\begin{pmatrix} \hat{a}_1^\dagger \\ \hat{a}_2^\dagger \end{pmatrix} = \exp(i\phi/2) \begin{pmatrix} C_1 & C_2 \\ -C_2^* & C_1^* \end{pmatrix} \begin{pmatrix} \hat{a}_3^\dagger \\ \hat{a}_4^\dagger \end{pmatrix} \quad , \tag{11}$$

where

$$\begin{aligned} C_1 &= t_1^* t_2^* \exp(i\phi/2) - r_1 r_2^* \exp\left(-\delta\vec{p}\cdot\vec{\nabla}_2\right)\exp\left(-\delta\vec{p}\cdot\vec{\nabla}_1\right)\exp(-i\phi/2) \quad, \\ C_2 &= -t_1^* r_2 \exp(i\phi/2)\exp\left(\delta\vec{p}\cdot\vec{\nabla}_2\right) - r_1 t_2 \exp\left(-\delta\vec{p}\cdot\vec{\nabla}_1\right)\exp(-i\phi/2) \end{aligned} \quad. \tag{12}$$

If we eliminate $BS_{out}$ then the above equations will be valid by assuming $t_2 = 1$, $r_2 = 0$.

Any two mode radiation state entering the two input ports of the interferometer can be described as a function of the operators $\hat{a}_1^\dagger$ and $\hat{a}_2^\dagger$ operating on the vacuum state. By using the transformations (11-12) one can transform this functions into a function of $\hat{a}_3^\dagger$ and $\hat{a}_4^\dagger$ operating on the vacuum state of the output modes. Using this straightforward method one can find the effect of the MZ interferometer on the transmitted radiation.

In [16] this method has been applied for coherent states. Here for simplicity we assume that one photon is entering into one input port of the MZ interferometer giving the input state as

$$a_1^\dagger |0\rangle_1 = \exp(i\phi/2)\left(C_1 a_3^\dagger |0\rangle_3 + C_2 a_4^\dagger |0\rangle_4\right) \quad, \tag{13}$$

where the subscript 1 refers to input port 1 and subscripts 3 and 4 refer to output ports 3 and 4, respectively. The wavefunction in output port 3 is orthogonal to that of 4 due to orthogonal space dependence. The probability for measurement of the photon in output port 3 is given by

$$|C_1|^2 = |t_1 t_2|^2 + |r_1 r_2|^2 - \left\{ t_1^* t_2^* r_1^* r_2 \exp(i\phi)\exp\left(\delta\vec{p}\cdot\vec{\nabla}_2\right)\exp\left(\delta\vec{p}\cdot\vec{\nabla}_1\right) + C.C. \right\} \quad, \tag{14}$$

and that in output port 4 is given by

$$|C_2|^2 = |t_1 r_2|^2 + |r_1 t_2|^2 + \left\{ t_1^* r_2 r_1^* t_2^* \exp(i\phi)\exp\left(\delta\vec{p}\cdot\vec{\nabla}_2\right)\exp\left(\delta\vec{p}\cdot\vec{\nabla}_1\right) + C.C. \right\} \quad. \tag{15}$$

The last term in Eq. (14) which is with opposite sign to the last term of Eq. (15) represents the interference term. We should take into account that the momentum translation operators with subscripts 1 and 2 operate on the macroscopic bodies $BS_{inp}$ and $BS_{out}$, respectively, due to the reflection of the photon from these BS's. The change in the BS's momentum wavefunctions due to the photon reflection is very small relative to their momentum uncertainty, as expressed for $BS_{inp}$ by Eq. (8) and by using a similar equation for $BS_{out}$. Therefore, in the ordinary treatment of MZ interferometer all the translational momentum operators can be neglected and Eqs. (14-15) are reduced to the conventional MZ analysis.

Let us see now what will be the effect of inserting the WW detector. Assuming for simplicity that the WW detector denoted as operator $\hat{O}_{WW}$ is interacting with the radiation reflected from $BS_{inp}$. Then Eq. (7) is exchanged into

$$\begin{pmatrix} \hat{a}_1^\dagger \\ \hat{a}_2^\dagger \end{pmatrix} = \begin{pmatrix} t_1^* & -r_1 \exp(-\delta\vec{p}\cdot\vec{\nabla}_1)\hat{O}_{WW} \\ r_1^* \exp(\delta\vec{p}\cdot\vec{\nabla}_1)\hat{O}_{WW}^\dagger & t_1 \end{pmatrix} \begin{pmatrix} \hat{b}_1^\dagger \\ \hat{b}_2^\dagger \end{pmatrix} \qquad . \tag{16}$$

Repeating again all the transformations with $\hat{O}_{WW}$ we find that in the expressions for $C_1$ and $C_2$ we should exchange $\exp(-\delta\vec{p}\cdot\vec{\nabla}_1)$ by

$$\exp(-\delta\vec{p}\cdot\vec{\nabla}_1) \to \exp(-\delta\vec{p}\cdot\vec{\nabla}_1)\hat{O}_{WW} \qquad . \tag{17}$$

Then in Eq. (12) terms which include $\hat{O}_{WW}$ are orthogonal to terms which do not include $\hat{O}_{WW}$. We find that Eqs. (14) and (15) are exchanged into

$$|C_1|^2 = |t_1 t_2|^2 + |r_1 r_2|^2 \quad ; \quad |C_2|^2 = |t_1 r_2|^2 + |r_1 t_2|^2 \qquad . \tag{18}$$

Due to the WW detector all interference terms are eliminated. By comparing Eqs. (14-15) with Eq. (18) we find that the WW detector has eliminated also the effects of the momentum translation operators. We find, therefore, that the WW detector has led to exchange of momentum [8,10] between the EM field and the entangled macroscopic BS's. However, such exchange of momentum is the <u>result</u> of the WW detector which is responsible to the which way effects [6,7,9,11]. Although many photons are involved in the interference effect this exchange of momentum has a negligible effect on the macroscopic BS's which have a large momentum uncertainty. Mometum transfer effects might be important only if micro-BS's are realizable. The intention of the present analysis is, however, to show that although small momentum exchanges occur in the WW experiments the cause of the WW effects is the introduction of orthogonality by WW detector [14] or equivalently by the complementarity of particle and wave properties [6,7,9,11].

## 4. Summary and conclusion

In the present work it has been shown in Section 2 that QM measurements are related to entanglement processes produced between the eigenstates of certain QM measurement operators and the macroscopic states of the apparatus of measurement as described in Eq. (1). In this expression only the degree of freedom of the macroscopic measuring device which is entangled with the QM system is taken into account while other degrees of freedom of the macroscopic system are disregarded. After producing such entanglement, the 'measurement' is made on the macroscopic entangled degree of freedom. Then using this information the total density operator is reduced to that of the QM system by tracing over the entangled states, excluding the macroscopic states. Two options for using such process as described by Eqs. (4) and (5). The method has been demonstrated for a simple Stern-Gerlach experiment.

In Section 3 the MZ transformations has been generalized so that they include momentum translation operators operating on the macroscopic BS's wavefunctions. This has led to final MZ transformation given by Eqs. (11-12). For simplicity of discussion we have assumed that one photon is inserted into one input port as described by Eq. (13). Then the probability for measuring the photon in output ports 3 and 4 are given by Eqs. (14) and (15), respectively. By inserting the WW detector in the reflected beam Eq. (7) is changed into Eq. (16) leading to orthogonality between the two routes of the interferometer excluding the interference terms. Equivalent results are obtained if we will put the WW detector in the beam transmitted from $BS_{inp}$. By comparing Eqs, (14-15) with (18) we find that the elimination of the interference terms by the WW leads also to exchange of momentum between the EM field and the BS's. However, we find that the effect of the momentum translation operators on the macroscopic wave function $BS_{inp}$ (and similarly on $BS_{out}$), as expressed by Eq. (8), are negligible, so that they cannot be regarded as the cause for the 'which way' effect, but only as a result of the collapse of the wavefunction from wave to particle property (i.e., complementarity principle [6,7,9,11]).

The idea that there is no reality before measurement [3] is adopted for cases in which we have superposition of states and interference effects. In cases for which superposition of states and interference effects are excluded reality is obtained already before measurement. This latter condition is essential for the definition of classical states including that of the classical world where all states are orthogonal.